\documentclass[]{aa}
\input{epsf}
\usepackage{amssymb}
\usepackage{natbib}
\usepackage{color}
\newcommand{\hMpc}{{\ifmmode{h^{-1}{\rm Mpc}}\else{$h^{-1}$Mpc }\fi}}
\newcommand{\hkpc}{{\ifmmode{h^{-1}{\rm kpc}}\else{$h^{-1}$kpc }\fi}}
\newcommand{\hMsun}{{\ifmmode{h^{-1}{\rm {M_{\odot}}}}\else{$h^{-1}{\rm{M_{\odot}}}$}\fi}}
\newcommand{\Msun}{{\ifmmode{{\rm {M_{\odot}}}}\else{${\rm{M_{\odot}}}$}\fi}}

\begin{document}

\title{Phase-space shapes of clusters and rich groups of galaxies} 
\author{Rados{\l}aw Wojtak}
\institute{Dark Cosmology Centre, Niels Bohr Institute, University of
  Copenhagen, Juliane Maries Vej 30, DK-2100 Copenhagen \O,
  Denmark}

\abstract{Clusters and groups of galaxies are highly aspherical, with shapes 
approximated by nearly prolate ellipsoids of revolution. An equally fundamental 
property is the shape of these objects in velocity space which is the anisotropy of 
the global velocity dispersion tensor. Although many studies address the problem 
of the shape in position space, there has been no attempt to measure shapes 
in velocity space.
}
{
Here we make use of kinematical data comprising $\sim 600$ nearby clusters and rich groups of galaxies from 
the Sloan Digital Sky Survey to place constraints on the phase-space shapes of these objects, i.e. their shapes 
in both position and velocity space.
}
{We show that the line of sight velocity dispersion normalised by a mass-dependent velocity scale 
correlates with the apparent elongation, with circular (elongated) clusters exhibiting an excessive (decremental) 
normalised velocity dispersion. This correlation holds for dynamically young or old clusters and, therefore, 
it originates from projecting their intrinsic phase-space shapes rather than from dynamical evolution. 
It signifies that clusters are preferentially prolate not only in position space, but also in velocity space. 
This property allows us to break the degeneracy between oblate and prolate models and thus to deproject the apparent elongations 
and the line of sight velocity dispersions obtaining constraints on the axial ratios of the ellipsoids approximating 
cluster shapes in 3D position or velocity space.
}
{
The distribution of the axial ratios in position space is found to be well 
approximated by a Gaussian with a mean, $\mu=0.66\pm0.01$, and a dispersion, $\sigma=0.07\pm0.008$. The velocity 
ellipsoids representing the shapes in velocity space are more spherical, with a mean axial ratio of $0.78\pm0.03$.
}
{The mean axial ratio of the velocity ellipsoids points to a highly anisotropic velocity distribution and, therefore, to 
a strong dependance of the observed velocity dispersions on the angle between the line of sight and the semi-principle 
axes of the clusters. This finding has important implications for mass measurements based on the line of sight velocity dispersion profiles 
in individual clusters. For typical axial ratios of the velocity ellipsoids in the analysed cluster sample, systematic errors on the mass estimates inferred 
from the line of sight velocity dispersions become comparable to statistical uncertainties for galaxy clusters 
with as few as $40$ spectroscopic redshifts.
}


\maketitle

\section{Introduction}

Clusters and rich groups of galaxies are the most massive gravitationally bound objects in the Universe. Many  
global properties such as luminosities, X-ray temperatures or velocity dispersions scale with the total mass 
of the host dark matter haloes \citep{Pop05} signifying that gravitational collapse is the primary physical mechanism 
shaping these objects. Scaling relations emerging from this process are of great importance for using clusters 
as cosmological probes. Conversion between various observable properties of galaxy clusters and their masses 
allows one to measure the mass function and thus to constrain cosmological parameters \citep[see e.g.][]{Roz10,Rap13}. 

There is no doubt that the halo mass is the main factor determining basic global properties of galaxy clusters. The corresponding 
gravitational potential well is deep enough to generate gravitational redshift of galaxies orbiting the cluster centres \citep{Woj11,Dom12}. 
The halo mass alone, however, is not sufficient for modelling a number of detailed features observed in galaxy clusters, 
from radial profiles of different observables to complex features such as morphology or substructures. Various models 
based on spherically symmetric density profiles of different mass components in galaxy clusters fit most observations 
of undisturbed clusters reasonably well \citep[see e.g.][]{Biv03,Vik06,New13}. They constitute a baseline for more accurate 
models taking into account asphericity of galaxy clusters.

Aspherical shapes of clusters and groups of galaxies is a well recognised property of these objects \citep[][and references therein]{Lim13}. The shapes 
can be approximated by triaxial ellipsoids which are preferentially prolate and have axial ratios of $\sim(0.5-0.8)$. They are commonly 
estimated by means of deprojecting the apparent elongations observed in optical or X-ray \citep[see e.g.][]{Bin82,Buo96,Wan08,Ser06}. 
Combining lensing, X-ray and Sunyaev-Zel'dovich observations of individual clusters often gives more accurate estimates and additional constraints 
on the shape of their host dark matter haloes \citep[see e.g.][]{Mor11,Mor12,Mor12a,Ser13} 

Asphericity of clusters and groups of galaxies has significant impact on precise measurements of the mass distribution in individual objects \citep{Cor09}. 
This is one of the main sources of systematic errors in mass measurements based on spherically symmetric models. It seems 
that further improvement in the cluster mass inference cannot avoid addressing this problem in a more rigorous way. Apart from 
this practical aspect of studies on the shapes of clusters and groups of galaxies, there is a question of its origin or relation to the cosmic web. 
Although the answer is not clear yet, a number of works show that galaxy groups and clusters are not randomly oriented, but rather they 
are aligned with the large-scale structures. This configuration is found in both observations \citep{Paz11} and cosmological simulations 
\citep{Fal02,Lib13}.

For groups and clusters observed optically, the shape describes the anisotropy of their galaxy distributions in 3D position 
space. It is commonly measured in terms of the axial ratios of the ellipsoids representing the moment of inertia tensor. Similar quantities 
can be defined in velocity space. This shape describes the anisotropy of galaxy distributions in 3D velocity 
space, i.e. the shape of the velocity ellipsoids representing the global velocity dispersion tensor. The shape in velocity space is 
not a derivative of the shape in position space, but it is complementary to it. When combined together, both quantities
\textit{give insight into the phase-space shape of galaxy distributions in groups or clusters, i.e. to what degree galaxy distributions 
are anisotropic in position and velocity space.} Although the literature is rich in measurements of the shapes in position 
space \citep[see e.g.][]{Bin82,Pli91,Bas00,Pli06,Wan08}, there is no estimate of the cluster shape in velocity space 
(the anisotropy of the global velocity dispersion ). In this work, we make use of kinematical data of clusters and rich groups 
of galaxies selected from the Sloan Digital Sky Survey (SDSS) \citep{Aih11} to measure the shapes in both position and 
velocity space. The results presented here are the first constraints on the full phase-space shapes of clusters and groups of galaxies.

The manuscript is organised as follows. In section 2, we describe the galaxy clusters and groups selected 
for this work as well as all observables used in the analysis. In section 3, we show how the measurements 
of the line of sight velocity dispersions combined with the apparent elongations can be used to discriminate 
between prolate and oblate models describing the phase-space shapes of groups and clusters. 
The main data analysis follows in sections 4 and 5, in which we measure the axial ratios of the ellipsoids representing 
the shapes of groups and clusters in position or velocity space. Section 6 includes the summary 
and conclusions.

\section{Data}
We make use of the catalogue of galaxy groups generated by \citet{Yan07}. 
The catalogue contains galaxy groups and members of the groups found with the hierarchical 
friends-of-friends (FOF) algorithm applied to the SDSS data. Its current up-to-date version is based 
on the 8th data release, DR8 \citep{Aih11}. For the purpose of our analysis, we select $574$ clusters and groups 
of galaxies containing at least $20$ members with spectroscopic redshifts.

The catalogue provides the total mass estimates of the groups based on abundance matching. The mass estimates are 
normalised to a fixed overdensity equal to $43\rho_{\rm c}$, where $\rho_{\rm c}$ is the critical density at 
the present time. Throughout the paper, we keep the same definition of the group mass based on 
the same overdensity parameter. The assumed overdensity threshold 
is approximately $2$ times smaller than typical virial overdensity in a $\Lambda$CDM cosmology, 
which is approximately equal to $100\rho_{\rm c}$ \citep[][]{Bry98}. This implies that the mass 
corresponding to the assumed overdensity is typically $30$ per cent larger than the commonly 
used virial mass.

The overdensity parameter assumed in the catalogue sets also a threshold for the projected phase-space 
density of galaxies selected as the group members (the density in the space of redshifts and the positions on the sky). 
This implies that typically $20$ per cent of the members populate a part of the infall zone extending 
to the projected distances which are $50$ per cent larger than the virial radius. A comparable fraction of cluster members 
happens to be found at distances up to $2$ virial radii due to a projection effect that affects all methods of the member selection \citep{Woj07}. 
Therefore, it seems that a lower phase-space density threshold adopted by \citet{Yan07}  
leads to a more balanced completeness along the sight line and in the plane of the sky than the 
commonly used virial overdensity, i.e. the maximum projected radii of the group members are comparable 
to the maximum line of sight distances resulting from the projection effect. 

Halo masses of the selected groups, as estimated from abundance matching, span the range from 
$2\times10^{13}\hMsun$ to $5\times10^{14}\hMsun$ ($95$ per cent range). The quartiles equal to 
$\log_{10}M_{\rm halo}[\hMsun]=13.7,\,14.0,\,14.2$. The sample comprises both clusters and rich groups of galaxies. 
For the sake of simplicity, hereinafter we refer to all selected objects as clusters.

The mass estimates from the catalogue are directly based on optical luminosities. Since the selection 
of cluster members does not explicitly depend on spherical symmetry, we expect that the luminosities and, 
therefore, the final mass estimates are not sensitive to the apparent shapes of the clusters (the same luminosities 
are expected for circular or elongated clusters with the same halo masses). In order to test this, we split the sample 
into equally-sized subsamples comprising circular and elongated clusters. If the mass estimates are independent of 
the elongations, one should expect the same mass distributions and mass-luminosity relations in both subsamples. 
Performing a K-S test we checked that the probability distributions underlying the two mass subsamples are the same 
(the null hypothesis that the distributions are the same cannot be rejected at the confidence level of $20$ per cent). 
We also checked that the luminosity-mass relations in the two cluster subsamples are fully consistent with each other. 
These two simple tests rule out any significant dependance of the mass estimates on the apparent shapes of the clusters.

For every cluster, we measure the projected velocity dispersion $\sigma_{\rm los}$ using a standard unbiased 
estimator for the variance, i.e.
\begin{equation}
\sigma_{\rm los}^{2}=\frac{1}{N-1}\sum_{i=1}^{N}(v_{i}-\hat{v})^{2},
\end{equation}
where $v_{i}$ are the line of sight velocities in the cluster rest frame and $\hat{v}$ is the mean. We also assign 
standard sampling errors given by $\sigma_{\rm los}/\sqrt{2(N-1)}$.

Using the positions of the cluster members on the sky, we calculate the shape tensor
\begin{equation}
T_{ij}=\sum_{k=1}^{N}r_{i,k}r_{j,k},
\end{equation}
where $(r_{1,k},r_{2,k})$ are the local Cartesian coordinates of $k$-th galaxy on the sky and $N$ is the number 
of members in a cluster. The coordinates are 
measured with respect to the cluster centre approximated by the mean position of all members. For calculating 
the mean position, we assign equal weights to all galaxies. We checked that weighting by luminosity does not change 
the final results of this work. By diagonalising the shape tensor, we calculate the semi-principle axes of the ellipses approximating 
the apparent shapes of the clusters. The semi-major and semi-minor axes are given by $a_{\rm 2D}=\sqrt{\lambda_{1}}$ and 
$c_{\rm 2D}=\sqrt{\lambda_{2}}$, where $\lambda_{1,2}$ are the eigenvalues of the shape tensor in increasing order. 
Finally, we quantify the shape of every cluster by computing the axial ratio (or the apparent elongation) $q_{2D}=c_{2D}/a_{2D}$. 
The errors on the apparent elongation are calculated by bootstrapping subsamples of galaxies from every cluster.

\section{Prolate or oblate ?}
In the first order approximation, the 3D shapes of clusters may be described by ellipsoids of revolution. Inferring 
the 3D shape ellipsoids from the observed axial ratios is difficult and always ends up with a degeneracy between prolate and oblate 
models. Here we show that combining measurements of the apparent axial ratios with the velocity dispersions allows one to break 
this degeneracy and to discriminate between prolate and oblate shape ellipsoids in a less ambiguous way.

The method relies on a very general argument that the velocity dispersion observed along the semi-major axis is larger than 
that along the semi-minor axis. This property may be easily shown using  the tensor virial theorem, although it is more 
general than all physical configurations permitted by this theorem. According to the theorem, the ratio of the corresponding diagonal 
elements of the kinetic and potential energy tensors calculated along the semi-principle axes are equal to $-2$. Since the element of 
the potential energy tensor associated with the semi-major axis is the largest, one expects that the sight line aligned with this axis maximises 
the observed velocity dispersion. This effect is well exhibited by simulated dark matter haloes as the alignment of the halo shape 
ellipsoids and the velocity ellipsoids -- geometrical representation of the velocity dispersion tensor \citep{Kas05,Woj13}.

The alignment of the semi-principle axes of the velocity dispersion tensor and the shape tensor leads to different observational signatures for populations of 
prolate or oblate galaxy clusters. The relative excess of the velocity dispersion is expected to occur in prolate clusters observed along their semi-major axis 
or in oblate clusters observed in the plane of two degenerate semi-major axes. Considering a sample of clusters observed along random 
sight lines, one can expect that the excess of the observed velocity dispersion should increase with the apparent elongation, when observing 
prolate clusters, or decrease with increasing elongation, when observing oblates clusters. Thus the problem of discriminating between 
prolate and oblate models is reduced to a test of correlation/anticorrelation between the line of sight velocity dispersion $\sigma_{\rm los}$ 
and the apparent elongation $q_{\rm 2D}$. It is worth emphasising that prolate/oblate models characterise the shapes in both position 
and velocity space.

\begin{figure}
\begin{center}
    \leavevmode
    \epsfxsize=8cm
    \epsfbox[75 75 560 410]{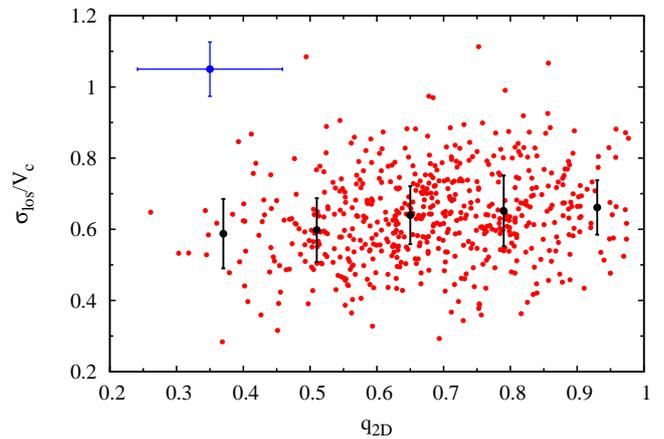}
\end{center}
\caption{Line of sight velocity dispersion in $574$ clusters and rich groups of galaxies as a function of the apparent elongation $q_{2D}$. The blue 
error bars show the mean $68$ per cent uncertainties of the measurements. The black points show the mean 
and the $68$ per cent scatter of the dispersions in $5$ bins of the elongation. The velocity dispersions are scaled by the characteristic 
velocity $V_{c}$ which is a measure of the gravitational potential depth of the host dark matter haloes. The figure shows that circular clusters 
exhibit higher normalised line of sight velocity dispersions than elongated ones.}
\label{s-e-nw}
\end{figure}

The velocity dispersion is primarily correlated with the halo mass. In order to remove the dependance on the halo mass, we normalise the dispersions 
by the characteristic velocity $V_{c}=\sqrt{GM_{\rm halo}/r_{\rm halo}}$, where $M_{\rm halo}=43\rho_{c}(4/3)\pi r_{\rm halo}^{3}$ is the halo mass from the group catalogue. 
The halo masses are estimated from abundance matching and, therefore, they are not explicitly related to the velocity dispersions. Thus the 
characteristic velocity $V_{c}$ sets the scale of the potential that is \textit{independent} of the observed velocities of galaxies in the clusters. 
This means that scaling by $V_{c}$ removes the dependance on mass (or the depth of the gravitational potential well), but preserves 
a key relation between excess/decrement of the velocity dispersion and the apparent elongation. 

Fig.~\ref{s-e-nw} shows the normalised line of sight velocity dispersions and the apparent elongations, $q_{2D}$, of the clusters. For the sake of readability, 
we do not plot the errors of every single measurement, but instead we show the mean $68$ per cent uncertainties resulting from averaging 
over all measurements (the blue error bars). The black points show the mean and the $68$ per cent scatter of the dispersions 
in $5$ bins of elongation. A visible trend of the mean increasing with $q_{2D}$ may point to a correlation between 
$\sigma_{\rm los}/V_{c}$ and $q_{2D}$. The measured scatter of data points exceeds the mean statistical uncertainty in $\sigma_{\rm los}/V_{c}$ 
by $17$ per cent. This signifies the existence of an intrinsic scatter in the normalised velocity dispersions. As we show later, this scatter can be ascribed 
to differences in the concentration of dark matter haloes. This scatter is comparable to the statistical uncertainties 
of the measurements and needs to be taken into account in quantifying the statistical significance of a plausible $\sigma_{\rm los}/V_{c}-q_{2D}$ correlation.

In order to estimate statistical significance of the correlation between the normalised line of sight velocity dispersion, $\sigma_{\rm los}/V_{c}$, 
and the apparent elongation, $q_{2D}$, we measure the slope of a linear relation between them. Linear fitting is 
carried out assuming the following likelihood function
\begin{eqnarray}
L(a,b,\sigma_{0}) & \propto & \prod \frac{1}{\sqrt{2\pi (\sigma_{i}^{2}+\sigma_{0}^{2})}} \nonumber \\
 & & \exp\Big(-\frac{[(\sigma_{los }/V_{c})_{i}-b-aq_{2D}]^{2}}{2(\sigma_{i}^{2}+\sigma_{0}^{2})}\Big),
\end{eqnarray}
where $\sigma_{i}$ is the statistical uncertainty on $(\sigma_{\rm los}/V_{c})_{i}$, $\sigma_{0}$ is the intrinsic scatter 
(assumed to be a free parameter), $a$ and $b$ are the slope and the intercept of a linear model, both free parameters.

\begin{figure}
\begin{center}
    \leavevmode
    \epsfxsize=8cm
    \epsfbox[75 75 560 410]{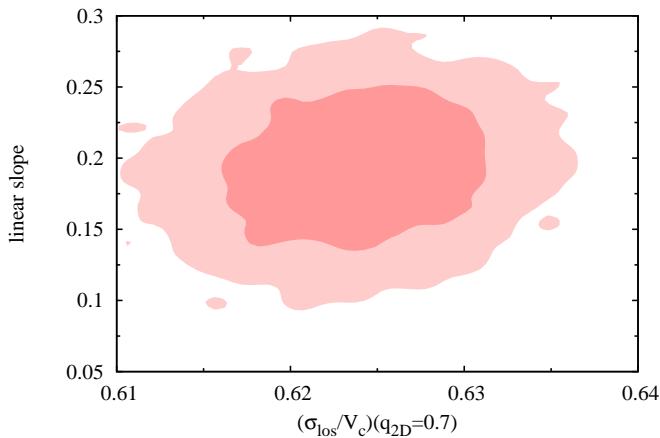}
\end{center}
\caption{Coefficients of a linear relation between the normalised line of sight velocity dispersion, 
$\sigma_{\rm los}/V_{c}$, and the apparent elongation, $q_{2D}$, of the clusters. The shaded  regions 
show the $68$ and $95$ per cent confidence contours of the posterior probability distribution marginalised 
over the intrinsic scatter in $\sigma_{\rm los}/V_{c}$. A positive slope signifies a correlation between the normalised 
line of sight velocity dispersion and the apparent elongation, consistent with prolate cluster shapes both 
in 3D position and 3D velocity space.}
\label{s-elong-MCMC}
\end{figure}

Constraints on the three parameters are obtained using an MCMC approach with the Metropolis-Hastings algorithm. In order to 
minimise a correlation between the slope and the intercept, we measure $q_{2D}$ with respect to the median equal to $0.7$. 
Fig.~\ref{s-elong-MCMC} shows the resulting constraints on the slope and the intercept marginalised over the intrinsic scatter. 
The $1\sigma$ intervals of the parameters are $0.19\pm0.04$ for the slope, $0.624\pm0.05$ for the intercept and $0.096\pm0.004$ 
for the intrinsic scatter. The analysis excludes a negative slope (anticorrelation) at the $>10^{-4}$ confidence level. We interpret this result as a 
strong observational signature of preferentially prolate shapes of the clusters.

The measurement of the line of sight velocity dispersion depends in general on which galaxies are selected as the cluster members. 
Since galaxy clusters tend to be aligned with the large-scale structures, one may suspect that the bias in the line of sight velocity dispersion, 
due to the background/foreground galaxies misidentified as the cluster members, may depend on the cluster orientation with respect to the line of sight. 
The presence of this kind of bias can be easily verified by measuring the fraction of blue galaxies as a function of the apparent elongation: clusters 
aligned with the line of sight are expected to be more contaminated by blue galaxies from the large-scale structures than those oriented perpendicularly 
to the line of sight. Closer inspection of the photometric data does not reveal any trend between the fraction of blue galaxies identified by $g-r$ colour 
(see more details about the colour selection in subsection 3.1) and the apparent elongation, i.e. the mean fraction of blue galaxies 
is the same in all bins of the apparent elongations. Therefore, we conclude that it is unlikely that the observed relation between the normalised line of sight velocity dispersion and the apparent elongation could result from large-scale structures aligned with galaxy clusters.

\subsection{Effect of dynamical state}

Statistically significant correlation between the normalised line of sight velocity dispersion and the apparent shapes of the clusters 
is consistent with a simple geometrical model in which the correlation emerges from projecting the cluster shapes which are approximated 
by prolate ellipsoids of revolution. It was suggested in several studies that the observed correlation 
may also be a signature of dynamical evolution of clusters \citep[see e.g.][]{Tov09,Rag10}. This interpretation was supported by some 
observed trends of the apparent elongation and velocity dispersion with the fraction of early-type galaxies and compactness 
of galaxy groups: groups with late formation times (dynamically young) appear to have larger elongations, smaller dispersions, 
smaller concentrations of galaxies and smaller fractions of early-type galaxies. 

In order to check what is the impact of this scenario on the geometrical interpretation of the $\sigma_{\rm los}/V_{c}-q_{2D}$ correlation, 
we split the sample into dynamically young (later formation times) and dynamically old (earlier formation times) clusters. 
We use two independent diagnostics of dynamical state which are independent of kinematical data: the fraction of early-type galaxies 
and the luminosity gap between the first- and second-ranked galaxies in the $g$ band. Dynamically old (relaxed) clusters tend to have 
higher fractions of elliptical galaxies as a result of dynamically-driven transformation of morphological types \citep{But78,But84,Got05} 
and larger magnitude gaps of two brightest galaxies \citep[see e.g.][]{Dar07,Smi10}. Early and late type galaxies are identified 
by $g-r$ colour. We adopt the same boundary line separating red and blues galaxies as \citet{Yan07} \citep[see also][]{Bos08}, i.e.
\begin{equation}
(g-r)_{z=0.1}=0.76+0.15[\log_{10}(M_{\ast}h^2/\Msun)-10],
\end{equation}
where $M_{\ast}$ is the stellar mass estimated using the relation between the stellar mass-to-light ratio and colour of \citet{Bel03}.

\begin{figure}
\begin{center}
    \leavevmode
    \epsfxsize=8cm
    \epsfbox[75 75 560 410]{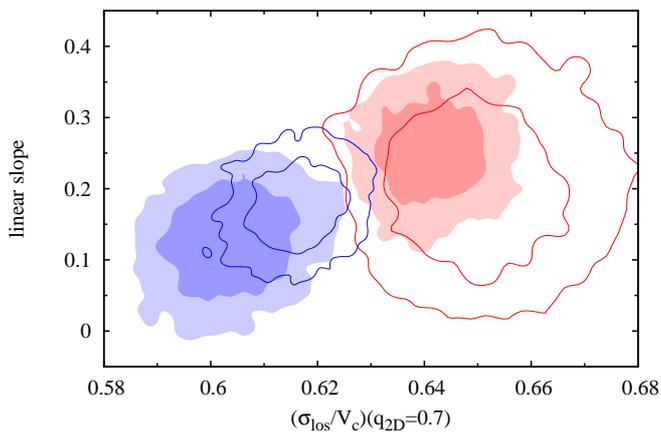}
\end{center}
\caption{Effect of dynamical state of clusters on the coefficients of a linear relation between the normalised line of sight 
velocity dispersion, $\sigma_{\rm los}/V_{c}$, and the apparent elongation, $q_{2D}$, inferred from the galaxy distribution on the sky. The red contours, 
with $(\sigma_{\rm los}/V_{c})(q_{2D}=0.7)\gtrsim 0.625$, show constraints for dynamically old clusters and the blue ones, 
with $(\sigma_{\rm los}/V_{c})(q_{2D}=0.7)\lesssim 0.625$, for dynamically young. The filled and empty 
contours correspond to the clusters selected by the fraction of red galaxies 
and by the magnitude gap of the two brightest galaxies, respectively. The contours show the $68$ and $95$ per cent 
confidence regions of the posterior probability distribution marginalised over the intrinsic scatter in $\sigma_{\rm los}/V_{c}$. 
Clusters at different dynamical states exhibit consistently a correlation between the normalised line of sight dispersion and 
the apparent elongation.
}
\label{s-elong-MCMC-dyn}
\end{figure}

We repeat fitting a linear model to the velocity dispersions and the apparent elongations of dynamically old and young clusters. The 
cluster sample is split into young and old systems using the median value of the fraction of early-type galaxies, 
$0.71$, and the magnitude gap of $0.8$. The old clusters appear to have larger $\sigma_{\rm los}/V_{c}$ than the 
young ones (see Fig.~\ref{s-elong-MCMC-dyn}). On the other hand, constraints on the slope of the 
$\sigma_{\rm los}/V_{c}-q_{2D}$ relation do not show any differences between the two cluster samples. 
Both the old and young clusters exhibit a strong correlation between $\sigma_{\rm los}/V_{c}$ 
and $q_{2D}$.

The obtained constraints on the linear coefficients signify that the effect of dynamical state manifests itself 
only as a \textit{modulation} of the $\sigma_{\rm los}/V_{c}-q_{2D}$ correlation, but not as the correlation itself. 
We note that  a similar modulation of the $\sigma_{\rm los}/V_{c}-q_{2D}$ relation was shown by \citet{Rag10} 
who used a compactness of galaxy groups as a diagnostic of their dynamical state. As an additional consistency 
test, we also checked that the modulation does not occur when splitting the cluster sample by mass, i.e. constraints 
on the slope and the intercept in four equally-sized bins of the halo mass are statistically the same. This simple test 
shows that scaling by $V_{c}$ removes effectively the dependance on halo mass.

Having shown that the effect of dynamical evolution gives rise to the modulation in the intercept of the 
$\sigma_{\rm los}/V_{c}-q_{2D}$ relation, we conclude that the correlation itself originates 
from projecting the 3D shapes of clusters and it should be considered 
a strong observational signature of clusters being preferentially prolate objects. 
This conclusion is in line with a number of works studying the same problem with different methods 
and using different kinematical data of groups and clusters of galaxies 
\citep[see e.g.][and references therein]{Bas00,Paz06,Pli06,Ser06,Wan08,Lim13}. Compered to these results, however, our 
conclusions go further. Prolate shapes of galaxy clusters are expected \textit{not only in position space, but also 
in velocity space}, i.e. both the 3D shape and the velocity dispersion tensor can be represented geometrically 
by coaligned prolate ellipsoids of revolution. This feature reduces considerably the family of all possible models describing the 
phase-space shapes of the clusters. It allows us to break the prolate-oblate degeneracy and thus to deproject 
the apparent elongations and the line of sight velocity dispersions.

\section{Shape in position space}

Here we reconstruct the 3D shapes of the clusters by means of deprojecting the apparent elongations $q_{2D}$. 
Deprojection is unambiguous if one reduces the domain of all permitted solutions to prolate or oblate ellipsoids. 
Following our reasoning presented in the previous section, we assume prolateness of the shape ellipsoids, 
in consistency with the observed $\sigma_{\rm los}/V_{c}-q_{2D}$ correlation.

For a population of clusters with the same 3D shape, observed along random sight lines, the apparent 
elongation $q_{2D}$ (apparent axial ratio) occurs with the following probability
\begin{equation}
p_{1}(q_{2D})\textrm{d}q_{2D}=\frac{q_{3D}^{2}}{q_{2D}^{2}\sqrt{(1-q_{3D}^{2})(q_{2D}^{2}-q_{3D}^{2})}}\textrm{d}q_{2D},
\label{prob-shape}
\end{equation}
where $q_{3D}\leqslant 1$ is the true axial ratio of the shape ellipsoids in 3D space. 
More realistic models of the cluster shapes should account for an intrinsic distribution of $q_{3D}$ 
within the population. In this case, the probability distribution of the apparent elongation becomes 
a convolution of the probability (\ref{prob-shape}) with the intrinsic distribution of the axial ratio $q_{3D}$. We approximate 
the latter by a Gaussian with a mean $\mu_{q_{3D}}$ and a dispersion $\sigma_{q_{3D}}$. In order to conform 
to the allowed range of values, the Guassian is truncated at $q_{3D}=0$ and $q_{3D}=1$.

\begin{figure}
\begin{center}
    \leavevmode
    \epsfxsize=8cm
    \epsfbox[75 75 560 410]{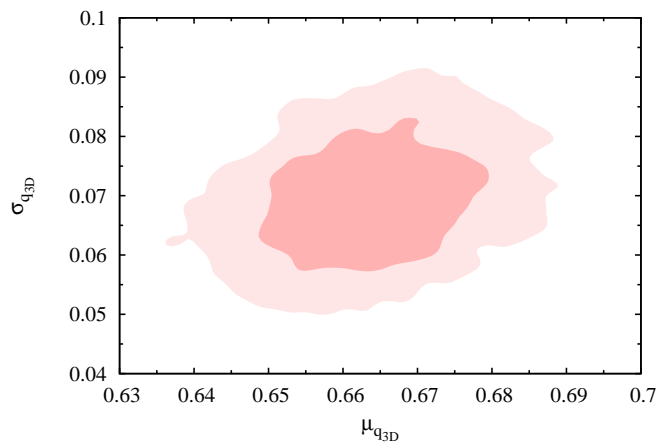}
\end{center}
\caption{Mean, $\mu_{q_{3D}}$, and dispersion, $\sigma_{q_{3D}}$, of the axial ratios of prolate ellipsoids approximating 
the 3D shapes of galaxy clusters. The contours show the $68$ and $95$ per cent confidence regions of the posterior probability.}
\label{3D-pos-MCMC}
\end{figure}

The errors on $q_{2D}$ are highly heterogenous and nearly comparable to the scatter resulting from 
the intrinsic distribution of $q_{3D}$. 
This feature needs to be taken into account in a robust  measurement of $\mu_{3D}$ and $\sigma_{3D}$. 
We incorporate the statistical uncertainties on $q_{2D}$ by convolving the model probability 
distribution $p_{\rm model}$ (the probability (\ref{prob-shape}) convolved with a Gaussian $G(\mu_{q_{3D}},\sigma_{q_{3D}})$) 
with the probability distribution $G_{\rm error\;i}$ describing constraints on the apparent elongation 
in consecutive clusters labeled by $i$. The corresponding likelihood function takes the following form
\begin{equation}
L \propto \prod_{i} [p_{\rm model}(\mu_{3D},\sigma_{3D})\ast G_{\rm error\;i}],
\label{like}
\end{equation}
where $\ast$ indicates a convolution. We approximate $G_{\rm error\;i}$ by a Gaussian with the maximum at 
$q_{2D\;i}$ and the dispersion equal to the error estimated from bootstrapping. The Gaussian is truncated at $q_{2D}=0$ 
and $q_{2D}=1$ and normalised to $1$ in the range $0\leqslant q_{2D}\leqslant 1$.

Fig.~\ref{3D-pos-MCMC} shows constraints on the mean, $\mu_{q_{3D}}$, and the dispersion, $\sigma_{q_{3D}}$, 
of the 3D axial ratios, $q_{3D}$, of the shape ellipsoids. The analysis was carried out using the MCMC technique. 
The best-fit values of the parameters are $\mu_{q_{3D}}=0.664\pm0.010$ and $\sigma_{q_{3D}}=0.069_{-0.007}^{+0.008}$. 
We checked that the constraints do not change when using subsamples of dynamically old or young clusters.

Our constraints on the 3D shapes of galaxy clusters are consistent with an independent measurement 
based on an earlier version of the catalogue by \citet{Yan07}, generated from SDSS DR4. A Monte Carlo 
deprojection of the apparent shapes carried out by \citet{Wan08} yield the mean axial ratio of around $0.6$ 
for groups with halo masses comparable to those from our sample and having at least $10$ members. Comparison with other results 
from the literature is less straightforward due to a number of differences between various measurements, e.g.  
the selection clusters or member galaxies. Neglecting all these differences, we find that most measurements 
of the shapes inferred from galaxy positions in clusters and massive groups point to a mean axial ratio 
of $\sim(0.5-0.6)$ and an intrinsic scatter of $\sim0.15$ \citep{Bin82,Pli91,deT95,Bas00}. Our results are fully 
compatible with these findings, although slightly larger mean and smaller scatter is preferred.

It is interesting to compare our results with the shapes of simulated dark matter haloes. Although the literature 
is rich in studies addressing this problem, such comparison is not straightforward, mostly 
because of different ways of measuring this property. 
However, selecting results which are based on a consistently similar working definition of the shape 
(given by the eigenvalues of the shape tensor calculated within the virial sphere of $\sim10^{14}\hMsun$ haloes), 
we find that the axial ratios (averaged over the minor-to-major and medium-to-major axis ratios) are typically 
in the range from $\sim0.6$ to $\sim0.7$ \citep[see e.g.][]{Bul02,Bai05,Kas05,All06,Bet07,Got07}. 
The measured axial ratios of the cluster shape ellipsoids, $q_{3D}\sim0.66\pm0.07$, happen to lie within 
this range.

\section{Shape in velocity space}

The distribution of the normalised line of sight velocity dispersion is partially determined by the shape 
of clusters in 3D velocity space, i.e. the anisotropy of the velocity dispersion tensor. 
Therefore, it can be used to place constraints on the anisotropy of the velocity dispersion quantified 
in terms of the ratio of the velocity dispersion along the semi-major axis, $\sigma_{\rm maj}$, to the velocity 
dispersion along the semi-minor axis, $\sigma_{\rm min}$. This novel idea is conceptually similar to a well-known 
technique of shape deprojection, but applied to the line of sight velocity dispersions instead of the apparent shapes on the sky.

Considering a population of clusters with the same anisotropy of the velocity dispersion tensor 
(the same axis ratios $q_{v}=\sigma_{\rm min}/\sigma_{\rm maj}$ of the velocity ellipsoids), 
one can show that the normalised velocity dispersion $\hat{\sigma}_{\rm los}=\sigma_{\rm los}/V_{c}$ along a random 
sight line is observed with a probability given by
\begin{equation}
p_{2}(\hat{\sigma}_{\rm los})\textrm{d}\hat{\sigma}_{\rm los}=\frac{\hat{\sigma}_{\rm los}}{\sqrt{\hat{\sigma}_{\rm los}^{2}(1+2q_{v}^{2})-bq_{v}^{2}}}
\frac{1+2q_{v}^{2}}{\sqrt{b(1-q_{v}^{2})}}\textrm{d}\hat{\sigma}_{\rm los},
\label{prob-velocity}
\end{equation}
where $b$ is a bias parameter defining the ratio of the total velocity dispersion to the characteristic velocity $V_{c}$ in the following way:
\begin{equation}
\sigma_{\rm maj}^{2}+2\sigma_{\rm min}^{2}=bV_{c}^{2}.
\end{equation}
From an argument of self-similar dynamics of galaxy clusters, one expects that $b$ should take comparable values for all 
clusters in the sample. The exact value of the bias parameter relies on the definition of $V_{c}$ and on how this velocity scale is 
related to the cluster dynamics. Needles to say, one should expect $b$ to be around $1$ if $V_{c}$ is the circular velocity on 
the scale comparable to the virial overdensity. In particular, for a spherical cluster in dynamical equilibrium, with galaxies 
and dark matter distributed according to the NFW density profile \citep{Nav97} with the concentration  parameter 
$c=5$ \citep[for the overdensity $200\rho_{\rm c}$; see][]{Pra12} and with isotropic orbits, one expects $b\approx1.2$ when 
$V_{c}$ is the virial velocity. In general, the theoretical prediction for the bias is rather not precise enough. 
Therefore, $b$ is assumed to be a free parameter in the analysis.

\begin{figure}
\begin{center}
    \leavevmode
    \epsfxsize=8cm
    \epsfbox[75 75 560 410]{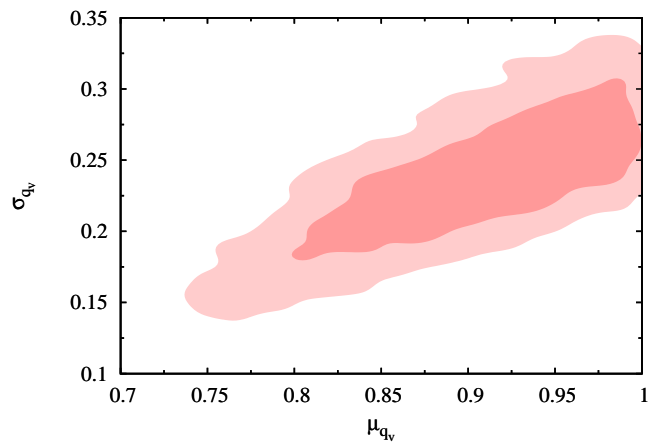}
\end{center}
\caption{Position of the maximum, $\mu_{q_{3D}}$, and dispersion, $\sigma_{q_{3D}}$, of a Gaussian 
approximating the distribution of the axial ratios $q_{v}=\sigma_{\rm min}/\sigma_{\rm maj}$ of prolate velocity ellipsoids representing 
the velocity dispersion tensor. The contours show the $68$ and $95$ per cent confidence regions of the posterior probability 
distribution marginalised over the bias parameter $b=\sigma_{\rm tot}^{2}/V_{c}^{2}$.}
\label{3D-vel-ms-MCMC}
\end{figure}

The remaining steps of the analysis are quite similar to the previous section. We assume that the intrinsic distribution of the 
axial ratio $q_{v}=\sigma_{\rm min}/\sigma_{\rm maj}$ is given by a Gaussian distribution with the maximum positioned at 
$\mu_{q_{v}}$ and the dispersion $\sigma_{q_{v}}$. The Gaussian 
is truncated at $q_{v}=0,1$ and normalised to $1$ in the range $0\leqslant q_{v} \leqslant 1$. 
Its convolution with the probability distribution (\ref{prob-velocity}) gives the final probability distribution of the observed $\sigma_{\rm los}/V_{c}$, 
i.e. $p_{\rm model}$ in the likelihood function (\ref{like}). Analogous to the shape deprojection, we use the likelihood 
given by (\ref{like}) with $G_{\rm error\; i}$ approximated by Gaussian. In this case however, the total 
dispersion $\delta_{i}$ of $G_{\rm error\; i}$ includes both the contributions from the statistical uncertainties on $\sigma_{\rm los}$ 
and the errors on $V_{c}$ resulting from a conversion between the luminosity and the cluster total mass:
\begin{equation}
\delta_{i}^{2}=\Big(\frac{\delta_{\sigma_{\rm los\;i}}}{V_{c\;i}}\Big)^{2}+
\Big(\frac{\sigma_{\rm los\;i}\delta_{V_{c\;i}}}{V_{c\;i}^{2}}\Big)^{2},
\end{equation}
where $\delta_{\sigma_{\rm los\;i}}$ is the sampling error of the dispersion estimator and $\delta_{V_{c\;i}}/V_{c\;i}$ is the relative 
accuracy in the estimate of $V_{c\;i}$ for $i$-th cluster. Typical relative error on the mass estimate from the mass-luminosity relation for galaxy 
clusters is estimated at $\delta M/M\approx 0.45$ \citep{Pop05}. Propagating this error to the characteristic velocity $V_{c}$ yields 
$\delta V_{c\;i}/V_{c\;i}=(1/3)(\delta_{M_{i}}/M_{i})\approx0.15$, which is the value adopted for the analysis.

As per the analysis of the cluster shapes in position space, we fit the model using the MCMC technique. Fig.~\ref{3D-vel-ms-MCMC} 
shows the resulting posterior probability distribution marginalised over the bias parameter. Constraints on $\mu_{q_{v}}$ 
and $\sigma_{q_{v}}$ are noticeably worse than for their counterparts in position space. This is not 
surprising, because this analysis attempts to recover 3D information from 1D data ($\sigma_{\rm los}$) which 
is a more complex problem than deprojecting 2D elongations $(q_{2D})$. Another feature displayed in Fig.~\ref{3D-vel-ms-MCMC} 
is a degeneracy between narrow Gaussians with smaller $\mu_{q_{v}}$ and wider Gaussians with larger $\mu_{q_{v}}$.

\begin{figure}
\begin{center}
    \leavevmode
    \epsfxsize=8cm
    \epsfbox[75 75 560 410]{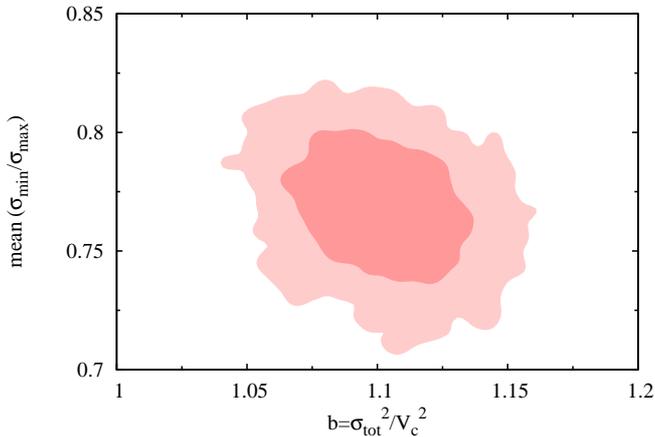}
\end{center}
\caption{Bias parameter, $b=\sigma_{\rm tot}^{2}/V_{c}^{2}$, and the mean 
anisotropy of the velocity dispersion tensor (axial ratio $\sigma_{\rm maj}/\sigma_{\rm min}$ of 
the velocity ellipsoids). The contours shows the $68$ and $95$ per cent confidence regions of the 
marginalised posterior probability distribution.}
\label{3D-vel-MCMC}
\end{figure}

Due to the fact that the distribution of $q_{v}$ is restricted to the range $0\leqslant q_{v} \leqslant 1$, and 
that the dispersion is comparable to the location of the maximum, $\mu_{q_{v}}$ \textit{is not} equal to the mean 
of the intrinsic distribution. Interestingly, the axis of degeneracy seen in Fig.~\ref{3D-vel-ms-MCMC} is approximately parallel 
to the lines of equal mean values. Therefore, it is tempting to marginalise the posterior probability distribution 
along the lines of constant mean values and thus to place tighter constraints on the mean axial ratio of the 
velocity ellipsoids. The resulting probability distribution is shown in Fig.~\ref{3D-vel-MCMC}. The mean axial 
ratio of the velocity ellipsoids equals to $q_{v}=0.78_{-0.03}^{+0.02}$ signifying \textit{highly anisotropic 
velocity distributions of galaxies in the selected groups and clusters.} This measurement happens to be 
compatible with typical values of the anisotropy of the global velocity dispersion tensor found in simulated 
dark matter haloes \citep{Kas05,Woj13}. We note that this approach to quantifying the degree of symmetry 
of the velocity distribution in galaxy clusters relies on more general reasoning than commonly used 
methods based on measuring the ratio of the radial-to-tangential velocity dispersions \citep[see e.g.][]{Biv04,Hwa08,Woj10}. 
Reasoning about isotropy or anisotropy based on the radial-to-tangential velocity dispersion requires an assumption 
of spherical symmetry which is in contradiction with the observed shapes of groups and clusters.

\begin{figure}
\begin{center}
    \leavevmode
    \epsfxsize=8cm
    \epsfbox[75 75 560 410]{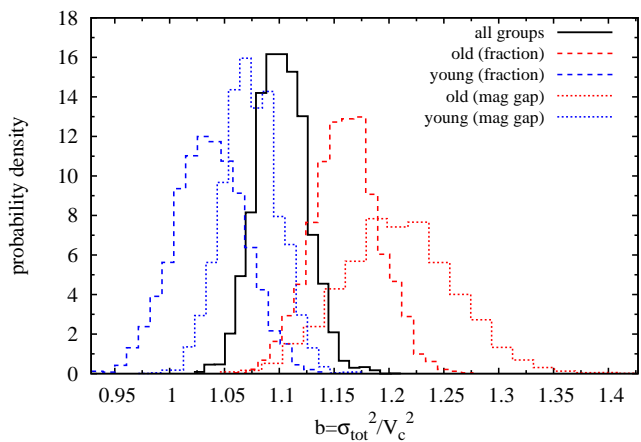}
\end{center}
\caption{Bias parameter $b=\sigma_{\rm tot}^{2}/V_{c}^{2}$ measured in dynamically old (red), 
young (blue) and all (black) clusters. The dashed and dotted lines show the marginalised posterior 
probability distributions for dynamical states differentiated by the fraction of red galaxies or by the magnitude gap of two brightest galaxies, 
respectively. Dynamically old clusters tend to have deeper gravitational potential well 
(larger bias parameter) than dynamically young clusters with the same halo masses.
}
\label{bias}
\end{figure}

We repeat the above analysis using subsamples of dynamically young and old clusters. 
The only unambiguous and statistically significant difference between 
the two subsamples occurs in constraints on the bias parameter (see Fig.~\ref{bias}). Dynamically old clusters appear to have 
higher bias parameters ($b=1.18_{-0.04}^{0.01}$ for clusters dominated by red galaxies and 
$b=1.24_{-0.08}^{0.02}$ for clusters with large magnitude gaps) than young ones ($b=1.04\pm0.03$ 
for clusters less dominated by red galaxies and $b=1.07_{-0.02}^{+0.03}$ for clusters with small magnitude 
gaps). This trend in the bias parameter results mainly from not taking into account the concentration 
of the dark matter distribution in the definition of the characteristic velocity $V_{c}$. As a consequence of this, $V_{c}$ is relatively 
smaller or bigger than the gravitational potential depth of dynamically young (with low-concentration haloes) 
or old (with high-concentration haloes) clusters, respectively. For a self-similar phase-space distribution 
of galaxies in clusters, this leads to a respectively higher or lower bias parameter, in proportion to the gravitational 
potential depths in the two cluster subsamples. A simple estimate of this effect can be made 
using an NFW gravitational potential \citep{Nav97,Lok01}. The mean concentration parameters $c$ 
of $\sim10^{14}\hMsun$ haloes lying above or below the mean mass-concentration relation 
are estimated at $c\approx7.6$ and $c\approx4.7$, respectively \citep[with the virial overdensity $95\rho_{c}$; see][]{Mac08}. 
This leads to the relative difference between the potential depths of $17$ per cent, in agreement with the relative 
change of the bias parameter between dynamically old and young clusters. In general, constraints on the bias parameter 
lie very close to a simple theoretical estimate based on an isotropic Jeans equation, $b=1.2$. The bias 
parameter inferred from an analysis combining both the old and young clusters amounts to $b=1.1_{-0.03}^{+0.01}$.

\section{Summary and conclusions}

We analysed galaxy kinematics in 574 clusters and rich groups of galaxies selected from the SDSS. We 
showed that the line of sight velocity dispersion normalised by a mass-dependent velocity scale  correlates 
with the apparent elongation of the clusters (higher velocity dispersion for more circular clusters). The 
correlation holds for dynamically young and old clusters (selected by the magnitude gap between two brightest 
galaxies or by the fraction of red galaxies) which signifies that it originates from the intrinsic anisotropy of galaxy 
distributions in phase space rather than from dynamical evolution. The sign of the correlation is consistent 
with prolate shapes of the clusters in both position and velocity space, i.e. the shape and velocity ellipsoids 
representing the moment of inertia and the velocity dispersion tensor are approximated by coaligned prolate ellipsoids 
of revolution. This property allowed us to deproject the apparent elongations and the line of sight velocity dispersion 
and thus to place constraints on the axial ratio of the shape and velocity ellipsoids.

Groups and clusters appear to be highly aspherical in both position and velocity space. Velocity 
ellipsoids with a mean axial ratio of $0.78$ are more spherical than shape ellipsoids which have typical axial 
ratios of $0.66$. Constraints on the shape ellipsoids complement consistently a number of previous 
measurements based on similar studies of galaxy positions in groups and clusters 
\citep[see e.g.][]{Bin82,Pli91,deT95,Bas00,Wan08}.

In our work we present the first estimate of asphericity of galaxy velocity distributions in clusters and rich groups, 
i.e. the degree of departure from spherical symmetry. Both the measurement and the reasoning are more general than 
a common approach based on estimating the so-called $\beta=1-\sigma_{\theta}^{2}/\sigma_{r}^{2}$ \citep[see e.g.][]{Biv04,Hwa08,Woj10}, 
which is incapable of assessing the symmetry of velocity field, because it implicitly assumes spherical symmetry. 

The constrains on the shape of velocity ellipsoids point to \textit{highly anisotropic velocity distributions}. 
When expressed in the same scale as the $\beta$ parameter, the mean anisotropy of the global velocity dispersion 
tensor equals to $1-\sigma_{\rm min}^{2}/\sigma_{\rm maj}^{2}\approx 0.4$. The total velocity dispersion 
normalised by the circular velocity based on the overdensity mass is higher for dynamically old clusters than for 
dynamically young ones. We ascribe this effect to differences in the concentration of the mass distribution between 
clusters at different states of dynamical evolution.

Highly anisotropic velocity distributions of galaxies in clusters and rich groups have important impacts on inferring 
halo masses or mass profiles from the observed line of sight velocity dispersions. Due to the unknown 
angle between the sight line and the principal axes of the velocity ellipsoids, the anisotropy becomes inevitably 
a major source of systematic errors. For the mean anisotropy of the velocity dispersion tensors, $\sigma_{\rm min}/\sigma_{\rm maj}=0.78$, 
\textit{the expected systematic error describing the deviation of the observed line of sight dispersion from the dispersion 
averaged over random sight lines amounts to $11$ per cent. 
This error becomes comparable to the statistical uncertainty on the line of sight velocity dispersion for clusters with as few 
as $40$ spectroscopic redshifts.} For rich clusters with more than $\sim100$ redshifts, the unknown projection of the velocity 
ellipsoid is expected to be one of the major sources of systematic errors in an analysis assuming spherical symmetry. 
This questions the relevance of increasing the complexity 
of dynamical models based on spherically symmetric phase-space density models. We expect that further 
improvement in constraining the mass profiles from kinematics of galaxies in clusters or rich groups should rely on 
proper modelling of asphericity rather then expanding degrees of freedom of spherical models. It is also worth noting that 
the number of spectroscopic redshifts available for some rich galaxy clusters, such as Abell 1689 \citep{Lem09}, is large enough to constrain the 
phase-space shapes of individual galaxy clusters. Our finding 
is also a caution for studying kinematics in such objects as dwarf spheroidals, for which the apparent 
elongations may likely be a signature of anisotropic velocity distributions and thus point to a limited 
applicability of spherically symmetric dynamical models \citep[see also][]{Kow13}.

\begin{acknowledgements}
The author is grateful to William Watson for critical reading of the manuscript. He also thanks 
Gary A. Mamon, Anja von der Linden and Andrea Biviano for fruitful discussions, and the anonymous 
referee for insightful comments. The Dark Cosmology Centre is funded by the Danish National Research Foundation.
\end{acknowledgements}

\bibliography{master}

\end{document}